\documentclass[a4paper,11pt]{article}
\usepackage{geometry}\usepackage[utf8]{inputenc}\usepackage{t1enc}
\usepackage{amsmath}\usepackage{amssymb}
\usepackage{graphicx}
\usepackage{xcolor}
\usepackage{hyperref}

\title{Supersymmetric Chern-Simons Theory in Presence of  a Boundary in the Light-Like Direction}
\author{Ji\v{r}\'{\i} Voh\'{a}nka\footnote{vohanka@physics.muni.cz} {} and Mir Faizal\footnote{f2mir@uwaterloo.ca}
\\ \\
\footnotemark[1] Department of Theoretical Physics and Astrophysics, Masaryk University,\\
Kotl\'{a}\v{r}sk\'{a} 267/2, 611 37 Brno, Czech Republic \\ 
\footnotemark[2] Department of Physics and Astronomy,   University of Waterloo, \\
Waterloo, Ontario N2L 3G1, Canada 
}
\date{}
\begin{document}

\newcommand{\fsl}[1]{#1\kern-0.50em/}
\newcommand{\fsnabla}{\nabla\kern-0.75em/}
\newcommand{\di}{\text{d}}
\newcommand{\pxu}[1]{\partial_\times{}^{#1}}
\newcommand{\pxd}[1]{\partial_{\times{#1}}}
\newcommand{\pxm}{\partial_{\times-}}
\newcommand{\gp}{\gamma_+}
\newcommand{\gpp}{\gamma_{++}}
\renewcommand{\wp}{w_+}
\newcommand{\fpp}{f_{++}}
\newcommand{\gx}{\gamma_\times}
\newcommand{\gxp}{\gamma_{\times+}}
\newcommand{\gxx}{\gamma_{\times\times}}
\newcommand{\wm}{w_-}
\newcommand{\fmp}{f_{+-}}
\newcommand{\fmm}{f_{--}}
\newcommand{\gm}{\gamma_-}
\newcommand{\gmp}{\gamma_{+-}}
\newcommand{\gmm}{\gamma_{--}}
\newcommand{\pp}{d_+}
\newcommand{\ppp}{\partial_{++}}
\newcommand{\pmp}{\partial_{+-}}
\newcommand{\pmm}{\partial_{--}}
\newcommand{\pxx}{\tfrac{\Box}{\partial_{++}}}
\newcommand{\exl}[2]{\underset{\text{#1}}{#2}}
\newcommand{\exr}[1]{\text{#1}}
\newcommand{\li}[1]{k_{#1}}
\newcommand{\lp}{\li{+}}
\newcommand{\lm}{\li{-}}
\newcommand{\lpp}{\li{++}}
\newcommand{\lmp}{\li{+-}}
\newcommand{\lmm}{\li{--}}
\newcommand{\lx}{\li{\times}}
\newcommand{\lxp}{\li{\times +}}
\newcommand{\lai}[1]{k^{(s)}_{#1}}
\newcommand{\lap}{\lai{+}}
\newcommand{\lam}{\lai{-}}
\newcommand{\lapp}{\lai{++}}
\newcommand{\lamp}{\lai{+-}}
\newcommand{\lamm}{\lai{--}}
\newcommand{\laxp}{\lai{\times +}}
\newcommand{\ti}[1]{v_{#1}}
\newcommand{\tp}{\ti{+}}
\newcommand{\tm}{\ti{-}}
\newcommand{\tpp}{\ti{++}}
\newcommand{\tmp}{\ti{+-}}
\newcommand{\tmm}{\ti{--}}
\newcommand{\tx}{\ti{\times}}
\newcommand{\txp}{\ti{\times +}}
\newcommand{\tai}[1]{v^{(s)}_{#1}}
\newcommand{\tap}{\tai{+}}
\newcommand{\tapp}{\tai{++}}
\newcommand{\tamp}{\tai{+-}}
\newcommand{\tamm}{\tai{--}}
\newcommand{\taxp}{\tai{\times +}}
\newcommand{\km}{\kappa_-}
\newcommand{\kx}{\kappa_\times}
\newcommand{\vm}{\nu_-}
\newcommand{\vx}{\nu_\times}
\newcommand{\Sxx}[2]{S^\text{#1}_\text{#2}}
\newcommand{\Sdx}[1]{S^\Delta_\text{#1}}
\newcommand{\pSxx}[2]{S^{\prime\text{#1}}_\text{#2}}
\newcommand{\pSdx}[1]{S^{\prime\Delta}_\text{#1}}

\maketitle

\begin{abstract}
In this paper, we will analyze a three dimensional supersymmetric Chern-Simons theory on a manifold with a boundary.
The boundary we will consider in this paper will be defined by   $n\cdot x=0$, where
$n$ is a light-like vector. It will be demonstrated that this 
boundary is preserved under the action of the $SIM(1)$ subgroup of
the Lorentz group. Furthermore, the 
  presence of this boundary will break half of the supersymmetry of the original theory.
  As the original Chern-Simons theory 
had $\mathcal{N} =1$ supersymmetry in absence of a boundary, it will only have 
 $\mathcal{N}=1/2$ supersymmetry in presence of this boundary.  
 We will also observe that the Chern-Simons 
 theory can be made gauge invariant by introducing new degrees of freedom 
 on the boundary. The gauge transformation of these new degrees of freedom will exactly cancel the 
 boundary term obtained from the gauge transformation of the Chern-Simons theory. 
\end{abstract}

\section{Introduction}

It is known that the action for  most renormalizable quantum field theories is at most quadratic in derivatives. 
This also includes the supersymmetric quantum field theories. So, the supersymmetric transformation of the action 
for these theories is expected to produce a total derivative term, apart from 
the bulk term. The bulk term vanishes due to the equations of motion, and the 
 total derivative term   vanishes in absence of a boundary. 
However, if a boundary is present,  this total derivative term will give rise to  a boundary 
 contribution. Thus, the presence of a boundary is expected to break  the supersymmetry of the theory. 
In fact,  as the presence of a boundary    breaks 
the translational invariance of the theory, and the translation invariance of the theory is related to 
the invariance of the theory under supersymmetry \cite{1}, it is expected that the supersymmetry will be broken 
due to the presence of a boundary. 
It is possible to impose suitable boundary conditions such that the supersymmetry of the theory will not be broken 
\cite{a}-\cite{b}.
These  boundary conditions are imposed on the Euler-Lagrange field equations. The surface terms vanish on-shell, after these boundary conditions are imposed, 
and this preserves the     on-shell supersymmetry of the theory. However, the boundary conditions imposed on the Euler-Lagrange field equations can not be 
used to preserve the off-shell supersymmetry of the theory.
It may be noted that various    
  boundary conditions  for supersymmetric theories have been analyzed 
\cite{c}-\cite{g}.  
The  path integral formalism is used to quantize most 
  supersymmetric theories, and this formalism uses 
off-shell fields. So, it is important 
 to have a formalism which preserves the  off-shell supersymmetry  in presence of a boundary. 

 Such a formalism has been constructed, and in this formalism the  half of the original supersymmetry is preserved 
 off-shell. This formalism is based on modifying the original action by adding a boundary contribution to it. 
 The supersymmetric variation of the boundary contribution exactly cancels the supersymmetric variation of the 
 bulk theory. However, this can only be done for half the supercharges of the original theory. Hence, only half the 
 supersymmetry of the original theory is preserved. This formalism has been used for 
  three dimensional theories with 
$\mathcal{N} =1 $ supersymmetry  \cite{1z}-\cite{9}.
This three dimensional formalism has   been used for analyzing a system of 
multiple M2-branes ending on a 
M5-brane \cite{4z}.  The action for multiple M2-branes is
dual to the supergravity on $AdS_4 \times S_7$, and the $OSp(8|4)$ 
symmetry of the eleven dimensional supergravity on $AdS_4 \times S_7$ 
is realized as $\mathcal{N} = 8$ supersymmetry of this dual field theory. Furthermore, all the on-shell 
degrees of freedom of this theory are exhausted by the matter fields, so the gauge sector has to be described 
by a topological theory. It has been possible to construct such a theory which is a matter-Chern-Simons theory 
called the Aharony-Bergman-Jafferis-Maldacena (ABJM) theory \cite{abjm}-\cite{ab2}. Even though this theory only has $\mathcal{N} = 6$ supersymmetry,
it is expected that its 
supersymmetry can be enhanced 
to full $\mathcal{N} = 8$ supersymmetry  \cite{abjm2}-\cite{2abjm}. In fact, it coincides with a theory 
called the Bagger–Lambert–Gustavsson (BLG) theory for two M2-branes, and the BLG theory has $\mathcal{N} = 8$ supersymmetry  
\cite{1abcd}-\cite{3abcd}. 

The   action for the matter sector of this theory  
 is gauge invariant even in presence of a boundary.
The action for the gauge sector  of this theory is described by a Chern-Simons theory.
It is well known that the gauge transformation of a Chern-Simons theory produces a surface term. 
So, in the presence of a boundary the gauge transformation of the ABJM theory generates a nonvanishing boundary term. 
  However, it has been demonstrated that 
if new boundary degrees of freedom are   introduced, then gauge invariance  of the ABJM theory
can be restored even in presence  
of a boundary \cite{5}-\cite{6}. This is because the gauge transformation of these 
 new boundary degrees of freedom
exactly cancels the boundary contribution generated from the gauge transformation of the bulk action. 
It is important to study   the open multiple M2-brane action as it can be used 
to understand the physics of M5-branes. It may be noted that a 
 system of   M2-branes intersecting with M5-branes has been studied  
 using a  fuzzy funnel solution   \cite{d4}-\cite{d5}.
A system of multiple M2-branes ending on two M9-branes is expected to generate
$E_8 \times E_8$ symmetry from the gravitational anomaly  \cite{hwsetup}-\cite{hwsetup1}. 
The BLG theory has been  used to study novel quantum geometry 
on the M5-brane world-volume by analyzing M2-branes ending on 
M5-branes with a constant $C$-field 
\cite{d12}, and 
 the BLG action with Nambu-Poisson 
$3$-bracket has been identified as the M5-brane action with a
large world-volume $C$-field \cite{M5BLG}. 
The boundary Chern-Simons theory has many 
other possible applications. It is possible for    D-branes to  
end  on other objects in string theory 
\cite{f1}-\cite{f2}. The Chern-Simons theory has   been used for analyzing a system of  
open strings ending on D-branes in 
the  the A-model topological string theory \cite{amodel}, and  the Holomorphic Chern–Simons theory
has been used for  analyzing the B-model in the string theory \cite{bmodel}. Thus, it is important to 
study the Chern-Simons theory in presence of a boundary. 

It may be noted that the  gauge and supersymmetric 
invariance of   Chern-Simons-matter theories has already been studied for  
boundaries in the space-like direction \cite{1z}-\cite{4z}. However, such an analysis has not been performed for the
boundaries in the light-like direction. It might be possible to generalize the  
 formalism developed to preseve the gauge and supersymmetry of the Chern-Simons-matter theories in presence of a boundary along 
 a space-like direction to 
 preserve these symmetries for the Chern-Simons-matter theories in presence of a boundary along a light-like direction. Such an generalization will have 
 to take into consideration the crucial differences between boundaries along space-like and light-like directions. 
For a boundary along a space-like direction the pullback of the metric on
the boundary has rank two, but for  a  boundary along a light-like  direction the pullback of the metric has only rank one.
However, it is possible to use some additional structure that occurs only for a boundary along a light-like direction to construct a gauge and supersymmetric 
invariant Chern-Simons  theory in presence of a boundary  along the light-like direction. 
Even thought the Lorentz symmetry breaks for boundaries along both the space-like and light-like directions, the boundaries along the light-like direction 
preserves a sub-group of the Lorentz group. A boundary in a light-like direction,  preserves  the 
 $SIM(1)$ group of the spacetime. Thus, we will use the $SIM(1)$ superspace formalism   \cite{vf1}  
 to describe Chern-Simons theory in presence of a boundary 
 along a light-like direction.  It has been demonstrated that 
  half the supersymmetry of the Lorentz invariant theory can be retained, when the Lorentz symmetry is broken down to the $SIM(1)$ symmetry.   
  This is done without adding additional boundary terms to the original action. 
  The advantage of this $SIM(1)$ superspace formalism is that  
the one-loop effective action for various theories can be easily calculated using $SIM(1)$ superspace formalism. In fact, one-loop effective action 
for a Wess-Zumino model has been calculated  using $SIM(2)$ superspace formalism \cite{1v}. 
The calculation of such effective action for Chern-Simons theories even in presence a boundary is non-trivial using the methods developed 
for analysing space-like boundaries. It may be noted that we do not need to add additional boundary terms to preserve half the supersymmetry 
of a Chern-Simons theory in presence of a boundary along light-like direction, if we use the $SIM(1)$ superspace formalism. However, 
  for Chern-Simons theories, we have to  add additional boundary terms 
to preserve gauge symmetry, even in $SIM(1)$ superspace formalism.

\section{Chern-Simons Theory}

The gauge covariant derivatives
\begin{align}\label{l_def}
 \nabla_\alpha &= D_\alpha -i \Gamma_\alpha,
&
 \nabla_{\alpha\beta} &= \partial_{\alpha\beta} -i \Gamma_{\alpha\beta},
\end{align}
are expressed with the help of connections $\Gamma_\alpha$, $\Gamma_{\alpha\beta}$, where the spinor derivatives
satisfy
\begin{equation}
 \{ D_\alpha , D_\beta \} = -2 \partial_{\alpha\beta}.
\end{equation}
Sometimes we will use Latin uppercase indices $A,B,\ldots$ to represent both spinor $A=\alpha$ and
vector $A=\alpha\beta$ indices. In this notation we write \eqref{l_def} as
\begin{equation}\label{l_short}
 \nabla_A = \mathcal{D}_A -i \Gamma_A,
\end{equation}
where the derivatives are 
$\mathcal{D}_A=(\mathcal{D}_\alpha,\mathcal{D}_{\alpha\beta})=(D_\alpha,\partial_{\alpha\beta})$.
It is also useful to assign Grassmann parity to indices. A spinor index $A=\alpha$ will be Grassmann odd 
$\widetilde{A}=1$ and a vector index $A=\alpha\beta$ will be Grassmann even $\widetilde{A}=0$.

The (anti)commutators among gauge covariant derivatives are
\begin{align}\label{l_comm}
 \{\nabla_\alpha,\nabla_\beta\} &= -2\nabla_{\alpha\beta},
\qquad\qquad\qquad\quad
 [\nabla_\alpha,\nabla_{\beta\gamma}] = C_{\alpha(\beta} W_{\gamma)},
\nonumber\\
 [\nabla_{\alpha\beta},\nabla_{\gamma\delta}] &= -\frac{1}{2} C_{\alpha\gamma} F_{\beta\delta}
	-\frac{1}{2} C_{\alpha\delta} F_{\beta\gamma} 
	-\frac{1}{2} C_{\beta\delta} F_{\alpha\gamma}
	-\frac{1}{2} C_{\beta\gamma} F_{\alpha\delta},
\end{align}
where the field strengths are
\begin{align}\label{l_fld}
 W_\alpha &= -\frac{i}{2}D^\beta D_\alpha \Gamma_\beta - \frac{1}{2}[\Gamma^\beta,D_\beta\Gamma_\alpha]
  + \frac{i}{6}[\Gamma^\beta,\{\Gamma_\beta,\Gamma_\alpha\}],
&
 \nabla^\alpha W_{\alpha} &= 0,
\nonumber\\
 \Gamma_{\alpha\beta} &= -\frac{1}{2}\left( D_{(\alpha}\Gamma_{\beta)} -i \{\Gamma_\alpha,\Gamma_\beta\} \right),
&
 F_{\alpha\beta} &= \frac{1}{2}\nabla_{(\alpha} W_{\beta)}.
\end{align}

The connections are subject to the gauge transformation
\begin{equation}\label{l_fgt}
 \Gamma^{(K)}_A = e^{iK} \Gamma_A e^{-iK} + i e^{iK} ( \mathcal{D}_A e^{-iK} ),
\end{equation}
where $K$ is a scalar superfield.
The infinitesimal version of the above gauge transformations is
\begin{equation}\label{l_igt}
 \delta_g^{(K)} \Gamma_A = i [ K , \Gamma_A] + \mathcal{D}_A K.
\end{equation}

The $\mathcal{N}=1$ Chern-Simons action is 
\begin{equation}\label{l_cs}
 \Sxx{cs}{} [ \Gamma_A ] = \frac{k}{4\pi} \text{tr} \int\di^3x \di^2\theta \left( \Gamma^\alpha W_\alpha 
 - \frac{1}{6} \{ \Gamma^\alpha , \Gamma^\beta \} \Gamma_{\alpha\beta} \right),
\end{equation}
where $k$ is the level of the Chern-Simons theory.
The gauge transformations of the Chern-Simons theory give rise to a surface term. 
This surface term does not cause any troubles for a theory without a boundary, but breaks the gauge invariance if
a boundary is present.
The infinitesimal gauge transformation \eqref{l_igt} gives the surface term
\begin{equation}\label{l_igt_cs}
 \delta_g^{(K)} \Sxx{cs}{} [ \Gamma_A ] = \frac{k}{4\pi} \text{tr} \int\di^3x \di^2\theta \left( 
 D^\alpha \left( K W_\alpha - \frac{1}{3} K [ \Gamma^\beta ,\Gamma_{\alpha\beta} ] \right)
 - \frac{1}{6} \partial_{\alpha\beta} \left( K \{ \Gamma^\alpha , \Gamma^\beta \} \right)
 \right).
\end{equation}

\section{\texorpdfstring{$SIM(1)$}{SIM(1)} Supersymmetry}

The detailed derivation of the $SIM(1)$ supersymmetry can be found in \cite{vf1}, here we are going to just review some
basic facts that we are going to use in this paper. 
The $SIM(1)$ group is a subgroup of the Lorentz group that preserves a given light-like direction, this means
that there is a light-like vector $n$ which is preserved up to a rescaling by the action of the $SIM(1)$ group.
This condition can also be formulated in the language of the double cover group $SL(2,\mathbb{R})$ of 
the Lorentz group $SO_+(2,1)$. The light-like vector $n$ can be written as $n^{\alpha\beta}=\xi^\alpha \xi^\beta$, 
where the commuting spinor $\xi$ is determined uniquely up to a sign. The $SIM(1)$ group is a subgroup of $SO_+(2,1)$ that preserves the light-like vector $n$  up to a rescaling. This  corresponds to a subgroup of $SL(2,\mathbb{R})$ determined by the condition that $\xi$ is preserved up to a rescaling. In this paper, we will assume that $\xi$ and $n$ are chosen such that
their nonzero components are $\xi^+=1$ and $n^{++}=1$. 
The $SIM(1)$ transformation of a general spinor $\psi$ can be written as
\begin{align}\label{sptr}
 \begin{pmatrix} \psi'^+ \\ \psi'^- \end{pmatrix} &= 
 \begin{pmatrix} e^{-A}&-B \\ 0&e^{A} \end{pmatrix}
 \begin{pmatrix} \psi^+ \\ \psi^- \end{pmatrix}
&&\Leftrightarrow&
 \begin{pmatrix} \psi'_+ \\ \psi'_- \end{pmatrix} &= 
 \begin{pmatrix} e^{A}&0 \\ B&e^{-A} \end{pmatrix}
 \begin{pmatrix} \psi_+ \\ \psi_- \end{pmatrix},
\end{align}
where $A,B\in\mathbb{R}$. Notice, that instead of matrices from $SL(2,\mathbb{R})$ that we would use in the case of
the Lorentz group $SO_+(2,1)$, we use only subgroup of $SL(2,\mathbb{R})$ consisting of triangular matrices.

When the symmetry is reduced to the $SIM(1)$ subgroup of the Lorentz group, the space of spinors $\mathcal{S}$
do not constitute an irreducible representation. While the group $SL(2,\mathbb{R})$ is semisimple and all representations can be written as a sum of irreducible representations, the group $SIM(1)$ is solvable and not all representations can be written as a sum of irreducible representations. One   such representations is the one that we have on $\mathcal{S}$. The subspace $\mathcal{S}_{\text{invariant}}$ consisting of all spinors that satisfy the condition $\fsl{n}\psi = 0$ is irreducible, and it is the only irreducible subspace of $\mathcal{S}$.
However, we have an irreducible representation on the quotient space $\mathcal{S}_{\text{quotient}} = \mathcal{S}/\mathcal{S}_{\text{invariant}}$. 
In our choice of $n$, the space $\mathcal{S}_{\text{invariant}}$ consists of spinors for which the $\psi_+$
coordinate vanishes, the space $\mathcal{S}_{\text{invariant}}$ can be conveniently described if we choose in each equivalence class a representative which has the coordinate $\psi_-$ equal to zero.
The infinitesimal $SIM(1)$ transformations are
\begin{align}\label{sptri}
 \begin{pmatrix} 0 \\ \psi'_- \end{pmatrix} &= e^{-A} \begin{pmatrix} 0 \\ \psi_- \end{pmatrix},
&
 \left[ \begin{pmatrix} \psi'_+ \\ 0 \end{pmatrix} \right ] 
 &= e^{A} \left [ \begin{pmatrix} \psi_+ \\ 0 \end{pmatrix} \right].
\end{align}

The $SIM(1)$ supersymmetry is not the symmetry that we get directly from super-Poincare symmetry when the Lorentz symmetry is broken down to the $SIM(1)$ symmetry, we also have to break half of the supersymmetry of the theory. Thus, the $\mathcal{N}=1$ supersymmetry is reduced to $\mathcal{N}=1/2$ supersymmetry.
The part of supersymmetry that we keep corresponds to supersymmetry transformations generated by $\epsilon Q$,
with the infinitesimal anticommuting parameter satisfying $\fsl{n}\epsilon=0$.

The number of anticommuting coordinates parameterizing $SIM(1)$ superspace is half of the number of coordinates that parametrize the original $\mathcal{N}=1 $ superspace. 
Thus, the   $SIM(1)$ supersymmetry only contains a single  supercharge $S_+$, and there is only one anticommuting coordinate $\theta_-$ parameterizing $SIM(1)$ superspace. This  supercharge 
corresponds to the spinor derivative $\pp$.
Thus, the generator of the $SIM(1)$ supersymmetry and the corresponding 
  spinor derivative are given by 
\begin{align}
 S_+ &= \partial_+ + i\theta_- \partial_{++},
&
 d_+ &= \partial_+ - i\theta_- \partial_{++}.
\end{align}
They satisfy
\begin{align}
 \{S_+,S_+\} &= 2\partial_{++},
&
 \{S_+,d_+\} &= 0,
&
 \{d_+,d_+\} &= -2\partial_{++},
&
 \partial_+ \theta_- &= -i.
\end{align}
It may be noted that the anticommuting coordinate $\theta_-$ transforms under the $SIM(1)$ group as a spinor from $\mathcal{S}_{\text{invariant}}$. The spinor derivative  and the generator of the supersymmetry transform under 
the $SIM(1)$ group as spinors from $\mathcal{S}_{\text{quotient}}$.

\section{Boundary Supersymmetry}

In this section, we are going to investigate how the symmetry of a theory is reduced, if we assume that there is
a boundary consisting of points that satisfy the condition $n\cdot x = 0$.
We are going to show that the $SIM(1)$ supersymmetry arises naturally in this context \cite{vf1}. We will review this    discussion here, because
it demonstrates which surface terms are relevant for this boundary theory.

In our particular choice of $n$ the condition $n\cdot x = 0$ means that $x^{--}=0$. 
This set of points is preserved under the action of the $SIM(1)$ group, because the direction of $n$ is preserved. 
We can also perform shifts in $x^{++}$ and $x^{+-}$ directions that are generated by $P_{+-}$, $P_{--}$.
The shift in the $x^{--}$ direction does not preserve the boundary, thus the $P_{--}$ generator cannot be part of 
the symmetry group.

In order to determine which part of supersymmetry is preserved, we will assume that there is a scalar superfield 
$\Phi$ which is constrained by the condition that it vanishes on the boundary.
Such superfield may appear for example in a matter Chern-Simons  theory.
The amount of unbroken supersymmetry will follow from the requirement that the boundary condition
\begin{equation}\label{sim_bcond}
 \Phi \vert_{x^{--}=0} = 0,
\end{equation}
is invariant. The infinitesimal supersymmetry transformation changes this boundary condition as
\begin{multline}\label{sim_binv}
 \delta \Phi \vert_{x^{--}=0} = -(\epsilon^\alpha Q_\alpha \Phi) \vert_{x^{--}=0}
 \\
 = -\left[ 
  \epsilon^+(\partial_+ + \theta^+ \partial_{++} + \theta^- \partial_{+-}) \Phi
  + \epsilon^-(\partial_- + \theta^+ \partial_{+-} + \theta^- \partial_{--}) \Phi
  \right] \vert_{x^{--}=0}
 \\
 = -\epsilon^- ( \theta^- \partial_{--} \Phi ) \vert_{x^{--}=0}.
\end{multline}
This result clearly shows that the boundary condition \eqref{sim_bcond} is left unchanged only if the 
infinitesimal parameter $\epsilon$ satisfies the condition $\fsl{n}\epsilon=\epsilon^-=0$.
This is the same condition that we used to break down the $\mathcal{N}=1$ Lorentz supersymmetry down to
the $\mathcal{N}=1/2$ $SIM(1)$ supersymmetry.

The only difference between the symmetry that we have just described and the $SIM(1)$ supersymmetry from the
previous section is that $P_{--}$ is not part of the boundary supersymmetry algebra. This does not affect 
most of results that we have in the $SIM(1)$ supersymmetry.
It should also be clear that only the surface term which is a total $\pmm$ derivative will be relevant when
we will investigate the gauge invariance of the Chern-Simons action. 

\section{Chern-Simons Theory in the \texorpdfstring{$SIM(1)$}{SIM(1)} Superspace}

In order to write down the Chern-Simons theory in the $SIM(1)$ superspace we introduce the projections
\begin{equation}\label{s_def}
 \gamma_A = \Gamma_A \vert_{\theta_+=0},
\end{equation}
of connections $\Gamma_\alpha,\Gamma_{\alpha\beta}$ \cite{vf2}.
The projection $\vert_{\theta_+=0}$ removes the dependence on the anticommuting coordinate $\theta_+$, 
which does not parametrize the $SIM(1)$ superspace.
The gauge transformations are
\begin{equation}\label{s_fgt}
 \gamma_A^{(K)} = e^{ik} \gamma_A e^{-ik} + i e^{ik} ( \mathcal{D}_A e^{-ik} ),
\end{equation}
where $k$ is the projection $k=K\vert_{\theta_+=0}$ of the scalar superfield $K$ and
the derivatives are $\mathcal{D}_A=(\mathcal{D}_+,\mathcal{D}_{\alpha\beta})=(\pp,\partial_{\alpha\beta})$.
The infinitesimal version of the above gauge transformations is
\begin{equation}\label{s_igt}
 \delta_g^{(K)} \gamma_A = i [ k , \gamma_A] + \mathcal{D}_A k.
\end{equation}
The rules \eqref{s_fgt}, \eqref{s_igt} do not hold for $\gm$ because the coordinate $\theta^-$ was lost
when we made projection on to the $SIM(1)$ superspace, thus we do not have anything that would correspond to $D_-$.
Instead we define a projection $\km=(D_-K)\vert_{\theta_+=0}$ and instead of \eqref{s_igt} we have
the infinitesimal gauge transformation
\begin{equation}\label{s_igt_gm}
 \delta_g^{(K)} \gm = i [ k , \gm] + \km.
\end{equation}
The projections $\gp$, $\gm$, $\gpp$, $\gmp$, $\gmm$ provide all information that we need to describe the gauge theory.
There is only one constraint that they have to satisfy
\begin{equation}\label{s_constraint}
 \pp \gp = - \gpp + \frac{i}{2} \{ \gp , \gp \}.
\end{equation}

The infinitesimal $SIM(1)$ transformation change a spinor $\psi$, according to \eqref{sptr}, as
\begin{align}
 \delta_s \psi_+ &= A \psi_+ ,
&
 \delta_s \psi_- &= - A \psi_- + B \psi_+ .
\end{align}
In the case of superfields, the infinitesimal change is calculated by applying the above rules on each index that 
it carries plus the change caused by the shift in superspace coordinates. 

The Chern-Simons action can be written as a sum of the bulk part $\Sxx{cs}{bulk}$, which contains the part
that can be written as an integral over the whole space-time, and the boundary part $\Sxx{cs}{boundary}$,
which contains the part that can be written as a total $\pmm$ derivative and is nonvanishing only on the
boundary
\begin{equation}\label{s_cs}
 \Sxx{cs}{} [ \gamma_A ] = \Sxx{cs}{bulk} [ \gamma_A ] + \Sxx{cs}{boundary} [ \gamma_A ].
\end{equation}
We will assume that the space-time is infinite in directions tangent to the boundary, we will not keep track
of terms that can be written as total derivatives in these directions. 
The Chern-Simons theory on a manifold without a boundary in $SIM(1)$ superspace has already been discussed in \cite{vf2}. The action that has been obtained corresponds to $\Sxx{cs}{bulk}$, and it is given by 
\begin{equation}\label{s_bulk}
 \Sxx{cs}{bulk} [ \gamma_A ] = \frac{k}{4\pi} \text{tr} \int\di^3x \di\theta^+ \bigg(
  2 \gmp \wm + \gp \fmm - \gmm \wp -i \gp[\gmp,\gmm]
 \bigg).
\end{equation}
Because there was no boundary, the boundary part of the action $\Sxx{cs}{boundary}$ was not discussed.  
The the boundary action can be easily found if we look at the derivation of $\Sxx{cs}{bulk}$ in \cite{vf2}.
The only place where a total $\pmm$ derivative appeared was in the identity 
\begin{equation}
 \text{tr} \int\di^3x \di\theta^+ \big[ \pmm(\{\gp,\gp\}\gm) \big]
 = \text{tr} \int\di^3x \di\theta^+ \big[ 2\{\gp,\gm\}(\pmm\gp) + \{\gp,\gp\}(\pmm\gm) \big].
\end{equation}
The appropriate multiple of the left side gives the boundary part of the action, which is
\begin{equation}\label{s_boundary}
 \Sxx{cs}{boundary} [ \gamma_A ] = \frac{k}{4\pi} \text{tr} \int\di^3x \di\theta^+ \pmm \bigg(
 - \frac{i}{6} \{ \gp , \gp \} \gm
 \bigg).
\end{equation}
The projections $w_\alpha=W_\alpha\vert_{\theta_+=0}$ and $f_{\alpha\beta}=F_{\alpha\beta}\vert_{\theta_+=0}$
can be calculated as
\begin{align}\label{s_fld}
 \wp &= \pp \gmp - \pmp \gp -i [ \gp , \gmp ],
\nonumber\\
 \wm &= \frac{1}{2} \left( \pp \gmm - \pmm \gp -i [ \gp , \gmm ] \right),
\nonumber\\
 \fpp &= - \ppp \gmp + \pmp \gpp +i [ \gpp , \gmp ],
\nonumber\\
 \fmp &= \frac{1}{2} \left( - \ppp \gmm + \pmm \gpp +i [ \gpp , \gmm ] \right),
\nonumber\\
 \fmm &= - \pmp \gmm + \pmm \gmp +i [ \gmp , \gmm ].
\end{align}

We should also note that neither the bulk action $\Sxx{cs}{bulk}$ nor the boundary action $\Sxx{cs}{boundary}$ are 
separately $SIM(1)$ invariant, if a boundary is present. The $SIM(1)$ transformation of the bulk action results in a surface term
that has to be canceled by terms that we get from the $SIM(1)$ transformation of the boundary action. 
It can be shown that
\begin{align}\label{s_igs_cs}
 \delta_s \Sxx{cs}{bulk} [ \gamma_A ] = - \delta_s \Sxx{cs}{boundary} [ \gamma_A ] 
 = B\frac{k}{4\pi} \text{tr} \int\di^3x \di\theta^+ \pmm \left( \frac{i}{6} \gp \{ \gp , \gp \} \right).
\end{align}

\subsection{Infinitesimal Gauge Transformation}

Let us look at the surface term that we get as a result of an infinitesimal gauge transformation. 
We will keep track only of those terms that are important for our boundary theory, 
that is terms that contain $\pmm$ derivative.

The infinitesimal change of $\Sxx{cs}{bulk}$ was already calculated in \cite{vf2}
\begin{equation}\label{s_igt_bulk}
 \delta_g^{(K)} \Sxx{cs}{bulk} [ \gamma_A ] = \frac{k}{4\pi} \text{tr} \int\di^3x \di\theta^+ \pmm \big( 
 -k ( \pp \gmp ) + k ( \pmp \gp )
 \big),
\end{equation}
where we kept only the surface term which is a total $\pmm$ derivative.

The infinitesimal change of the boundary term can be calculated with the help of \eqref{s_igt} and \eqref{s_igt_gm}.
The result is
\begin{multline}\label{s_igt_1}
 \delta_g^{(K)} \Sxx{cs}{boundary} [ \gamma_A ] = \frac{k}{4\pi} \text{tr} \int\di^3x \di\theta^+ \pmm \bigg(
 - \frac{i}{6} \{ i [ k , \gp ] + \pp k , \gp \} \gm 
 \\
 - \frac{i}{6} \{ \gp , i [ k , \gp ] + \pp k \} \gm 
 - \frac{i}{6} \{ \gp , \gp \} ( i [ k , \gm ] + \km )
 \bigg).
\end{multline}
All terms with the commutator $i[k,\cdot]$ drop out of the calculation due to the cyclic property of the trace
and the super-Jacobi identity, moreover the term $\{ \pp k , \gp \} \gm$ appears twice, so the result could be
written as
\begin{equation}\label{s_igt_boundary}
 \delta_g^{(K)} \Sxx{cs}{boundary} [ \gamma_A ] =  \frac{k}{4\pi} \text{tr} \int\di^3x \di\theta^+ \pmm \bigg(
 - \frac{i}{3} (\pp k) \{ \gp , \gm \} - \frac{i}{6} \km \{ \gp , \gp \}
 \bigg).
\end{equation}

\subsection{Finite Gauge Transformation}

The gauge transformation generated by $K$ (or equivalently by $k$ and $\km$) changes the action by a boundary term, which we will be
denoted as $\Sdx{}$. We may write this as
\begin{equation}\label{s_fgt_cs}
 \Sxx{cs}{} [ \gamma_A^{(K)} ] = \Sxx{cs}{} [ \gamma_A ] + \Sdx{} [ \gamma_A ; k , \km ],
\end{equation}
where $\gamma_A^{(K)}$ are gauge transformed connections.
Note that the boundary term depends on both $\gamma_A$, $k$ and $\km$.
We will state the result for $\Sdx{}$ now, the proof will be provided later in this section.
The boundary contribution is
\begin{multline}\label{s_delta}
 \Sdx{} [ \gamma_A ; k , \km ] = \frac{k}{4\pi} \text{tr} \int\di^3x \di\theta^+ \pmm \Big[ 
 - \frac{i}{6} ( \gm + \lm ) \{ \gp + \lp , \gp + \lp \}
 + \frac{i}{6} \gm \{ \gp , \gp \}
 \\
 + \lp \gmp 
 - \gp \lmp
 + \int_0^1 \di s \left(  
    \left( \tfrac{\di}{\di s} \lap \right) \lamp 
  - \lap \left( \tfrac{\di}{\di s} \lamp \right)
 \right)
 \Big],
\end{multline}
with $\lai{A}$ and $\li{A}$ defined as
\begin{align}\label{s_k}
 \lai{A} &= i ( \mathcal{D}_A e^{-isk} ) e^{isk},
&
 \li{A} &= \lai{A} \vert_{s=1} = i ( \mathcal{D}_A e^{-ik} ) e^{ik}.
\end{align}
The superfield $\lm$ is not covered by the above definition for the same reasons as in \eqref{s_igt_gm}. 
Instead, it is defined as
\footnote{
This could be also written as
\begin{equation*}
 \lm = \int_0^1 \di s \left(
 e^{-isk} \km e^{isk}
 \right).
\end{equation*}
}
\begin{equation}\label{s_km}
 \lm = i ( D_- e^{-isK} ) e^{isK} \vert_{\theta_+=0}.
\end{equation}

The boundary contribution \eqref{s_delta} should depend only on the value of the group element $e^{iK}$, not on 
the value of superfield $K$ used to parametrize it. This is clearly true for terms that only contain $\li{A}$ and 
$\gamma_A$, because $\li{A}$ depend only on the group element $e^{ik}$ (and $e^{iK}$ in the case of $\lm$). 
The uniqueness of terms with $\lai{A}$ is not so clear, because there might be multiple choices of $k$ that give 
the same group element $e^{ik}$. We are going to show that for gauge groups that are simply connected and
have surjective exponential map the term with integral over $s$ is also well defined. 
The group $SU(N)$ is one of the groups for which these conditions are met. 
The $s$ integral can be understood as an integral along the curve 
$[0,1]\ni s \rightarrow g(s)=e^{isk}$ that connects the identity element with the element $e^{ik}$.
The integrand is a $1$-form
\begin{equation}\label{s_fgt_omega}
 \omega = \text{tr} \left[
   \di_g \big( (\pp g^{-1}) g \big ) \big( (\pmp g^{-1}) g \big ) 
 - \big( (\pp g^{-1}) g \big ) \di_g \big( (\pmp g^{-1}) g \big )
 \right],
\end{equation}
where $\di_g$ denotes the exterior derivative with respect to the group element $g$. It is trivial to show that 
$\di_g\omega=0$. The fact that the form $\omega$ is closed together with the assumption that the group is 
simply connected leads to the conclusion that the result of the integral is independent on the choice of a path
which we pick to connect the identity and the element $e^{ik}$.  
We need the surjectivity requirement of the exponential map to ensure that all group elements can be written 
as $e^{ik}$. It seems that this requirement would not be necessary, if we did not assume a particular parametrization and
expressed the result as a curve integral of $\omega$.

Although it is possible to calculate $\Sdx{}$ by applying gauge transformation \eqref{s_fgt}, we will use 
a different approach. We will show that \eqref{s_delta} is what we would get if we considered a finite gauge 
transformation as a series of infinitesimal ones. 

Consider a gauge transformation generated by $(1+\epsilon)K$, where $\epsilon$ is an infinitesimal parameter. 
According to \eqref{s_fgt_cs} we have
\begin{multline}\label{s_fgt_1}
 \Sxx{cs}{} [ \gamma_A^{(K+\epsilon K)} ] 
  = \Sxx{cs}{} [ \gamma_A ] + \Sdx{} [ \gamma_A ; k + \epsilon k , \km + \epsilon \km ]
 \\
  = \Sxx{cs}{} [ \gamma_A ] + \Sdx{} [ \gamma_A ; k , \km ] 
    + \epsilon \hat{\delta}_g^{(K)} \Sdx{} [ \gamma_A ; k , \km ],
\end{multline}
where $\epsilon\hat{\delta}_g^{(K)}$ is used to denote an infinitesimal transformation that changes $K$ to $(1+\epsilon)K$
but leaves $\gamma_A$ unchanged, i.e.
\begin{align}\label{fgt_k_s}
 \hat{\delta}_g^{(K)} k &= k,
&
 \hat{\delta}_g^{(K)} \km &= \km,
&
 \hat{\delta}_g^{(K)} \gamma_A &= 0.
\end{align}
We may also understand the gauge transformation generated by $(1+\epsilon)K$ as a composition of a finite gauge transformation with an infinitesimal gauge transformation. Here $K$ parameterizes  the finite gauge transformations, and  $\epsilon K$ parameterizes the infinitesimal gauge transformation. 
Thus, the alternative method to calculate the gauge transformation is
\begin{multline}\label{s_fgt_2}
 \Sxx{cs}{} [ \gamma_A^{(K+\epsilon K)} ] 
  = \Sxx{cs}{} [ \gamma_A^{(K)} ] + \epsilon \delta_g^{(K)} \Sxx{cs}{} [ \gamma_A^{(K)} ]
  \\
  = \Sxx{cs}{} [ \gamma_A ] + \Sdx{} [ \gamma_A ; k , \km ] 
    + \epsilon \delta_g^{(K)} \Sxx{cs}{} [ \gamma_A ] + \epsilon \delta_g^{(K)} \Sdx{} [ \gamma_A ; k , \km ].
\end{multline}
In this case $\delta_g^{(K)}$ changes $\gamma_A$ according to \eqref{s_igt}, \eqref{s_igt_gm} but 
does not affect $k$ and $\km$, so $\delta_g^{(K)}k=\delta_g^{(K)}\km=0$. Comparison of terms with $\epsilon$ in 
\eqref{s_fgt_1} and \eqref{s_fgt_2} gives the equation
\begin{equation}\label{s_fgt_eq}
 \hat{\delta}_g^{(K)} \Sdx{} [ \gamma_A ; k , \km ] - \delta_g^{(K)} \Sdx{} [ \gamma_A ; k , \km ]
 = \delta_g^{(K)} \Sxx{cs}{} [ \gamma_A ].
\end{equation}
In order to prove that $\Sdx{}$ is correct, we have to show that it satisfies the above equation and
the boundary condition $\Sdx{}[\gamma_A;0,0]=0$. The verification of the boundary condition is trivial, 
we have already calculated the right side of \eqref{s_fgt_eq} in \eqref{s_igt_bulk} and \eqref{s_igt_boundary}, 
what remains is to evaluate expressions on the left side. 
Before we proceed with it, we are going to derive a few useful identities. The first identity is
\begin{multline}\label{s_kda}
 \mathcal{D}_A \lai{B} - (-1)^{\widetilde{A}\widetilde{B}} \mathcal{D}_B \lai{A} = 
   i ( \mathcal{D}_A \mathcal{D}_B e^{-isk} ) e^{isk}
 + i (-1)^{\widetilde{A}\widetilde{B}} ( \mathcal{D}_B e^{-isk} ) (\mathcal{D}_A e^{isk} ) 
 \\
 - i (-1)^{\widetilde{A}\widetilde{B}} ( \mathcal{D}_B \mathcal{D}_A e^{-isk} ) e^{isk}
 - i ( \mathcal{D}_A e^{-isk} ) (\mathcal{D}_B e^{isk} )
 = \lai{[A,B]_\pm} - i [ \lai{A}, \lai{B} ]_\pm, 
\end{multline}
where we used that $\mathcal{D}_A e^{isk} = - e^{isk} (\mathcal{D}_A e^{-isk})  e^{isk}$.
The symbol $\lai{[A,B]_\pm}$ is used to denote
\begin{equation}\label{s_kdca}
 \lai{[A,B]_\pm} = i ( [ \mathcal{D}_A , \mathcal{D}_B ]_\pm e^{-isk} ) e^{isk}.
\end{equation}
For example, when we set $A=+$, $B=+$ we obtain the identity
\begin{equation}\label{s_kdpa}
 \pp \lap = \frac{1}{2}( \pp \lap + \pp \lap ) = - \lapp - \frac{i}{2} [\lap , \lap],
\end{equation}
where we used $\{\mathcal{D}_+,\mathcal{D}_+\}=\{\pp,\pp\}=-2\ppp=-2\mathcal{D}_{++}$.
Another identity that can be easily derived is  
\begin{equation}\label{s_kalpha}
 \frac{\di}{\di s} \lai{A} = \mathcal{D}_A k - i [ k , \lai{A} ].
\end{equation}
We can use the fact that $\hat{\delta}_g^{(K)} (sk) = s \frac{\di}{\di s} (sk)$ to find 
the infinitesimal transformation of $\lai{A}$
\begin{equation}\label{s_kia}
 \hat{\delta}_g^{(K)} \lai{A} 
 = s \frac{\di}{\di s} \lai{A}
 = s \left( \mathcal{D}_A k - i [ k , \lai{A} ] \right).
\end{equation}
If we set $s=1$ in \eqref{s_kda} we get
\begin{equation}\label{s_kd}
 \mathcal{D}_A \li{B} - (-1)^{\widetilde{A}\widetilde{B}} \mathcal{D}_B \li{A}
 = \li{[A,B]_\pm} - i [ \li{A}, \li{B} ]_\pm.
\end{equation}
Similar methods can be used to calculate the infinitesimal change $\hat{\delta}_g^{(K)}$ of $\li{A}$ and $\lm$
\begin{align}\label{s_ki}
 \hat{\delta}_g^{(K)} \li{A} &= \mathcal{D}_A k - i [ k , \li{A} ],
&
 \hat{\delta}_g^{(K)} \lm &= \km - i [ k , \lm ].
\end{align}

Now, we are ready to evaluate the expressions on the left side of \eqref{s_fgt_eq}. 
The first term in \eqref{s_delta} does not give any contribution because the infinitesimal change of combinations 
$\gamma_A + \li{A}$ is
\begin{equation}\label{s_fgt_3}
 \left( \hat{\delta}_g^{(K)} - \delta_g^{(K)} \right) ( \gamma_A + \li{A} ) = - i [ k , \gamma_A + \li{A} ],
\end{equation}
and the $-i[k,\cdot]$ commutators drop out because of the super-Jacobi identity and the cyclic property of the trace.
The other terms that are outside of the $s$-integral give
\begin{multline}\label{s_fgt_4}
 \left( \hat{\delta}_g^{(K)} - \delta_g^{(K)} \right) \left(
 \text{tr} \int\di^3x \di\theta^+ \pmm \Big[
 \lp \gmp - \gp \lmp + \frac{i}{6} \gm \{ \gp , \gp \}
 \Big]
 \right)
 \\
 = \text{tr} \int\di^3x \di\theta^+ \pmm \Big[
   ( \pp k ) \gmp - \lp ( \pmp k )
 + ( \pp k ) \lmp - \gp ( \pmp k )
 \\
 - \frac{i}{6} \km \{ \gp , \gp \} - \frac{i}{6} \gm \{ \pp k , \gp \} - \frac{i}{6} \gm \{ \gp , \pp k \}
 \Big].
\end{multline}
As before, all terms with $-i[k,\cdot]$ cancel among themselves and yield zero net contribution.
In the next step, we integrate the first four terms by parts to move the derivatives $\pp$, $\pmp$ in front of $k$ 
so they act on $\gamma_A$, $\li{A}$. 
The expression $k ( \pmp \lp - \pp \lmp )$ that we get from the second and the third term can be replaced with
$ik[\lp,\lmp]$ because of the identity \eqref{s_kd}. 
Thus, the result for the part that is outside of the $s$-integral is
\begin{multline}\label{s_fgt_lhs1}
 \text{tr} \int\di^3x \di\theta^+ \pmm \Big[
 - k ( \pp \gmp ) + k ( \pmp \gp )
 + i k [ \lp , \lmp ]
 \\
 - \frac{i}{6} \km \{ \gp , \gp \} - \frac{i}{3} (\pp k) \{ \gp , \gm \}
 \Big].
\end{multline}
The last piece which we need to evaluate on the left side of \eqref{s_fgt_eq} is the part of \eqref{s_delta} that is
inside the $s$-integral. 
The infinitesimal change $\delta_g^{(K)}$ does not give any contribution because $\delta_g^{(K)}\li{A}=0$.
The change $\hat{\delta}_g^{(K)}$ can be easily calculated if we write $\hat{\delta}_g^{(K)}\lai{A}$
as $s\frac{\di}{\di s}\lai{A}$. The result obtained  by following this procedure can be written as  
\begin{multline}\label{s_fgt_5}
 \hat{\delta}_g^{(K)} \left(
 \text{tr} \int\di^3x \di\theta^+ \pmm \int_0^1 \di s \left[ \left( 
   \tfrac{\di}{\di s} \lap \right) \lamp 
 - \lap \left( \tfrac{\di}{\di s} \lamp \right) 
 \right]
 \right)
 \\
 = \text{tr} \int\di^3x \di\theta^+ \pmm \int_0^1 \di s \left( 
 \tfrac{\di}{\di s} \left[
   s \left( \tfrac{\di}{\di s} \lap \right) \lamp
 - s \lap \left( \tfrac{\di}{\di s} \lamp \right)
 \right] 
 \right).
\end{multline}
The substitution for $\frac{\di}{\di s}\lai{A}$ according to \eqref{s_kalpha} and integration over $s$
gives
\begin{equation}\label{s_fgt_6}
 \text{tr} \int\di^3x \di\theta^+ \pmm \Big( 
 ( \pp k ) \lmp - i [ k , \lp ] \lmp - \lp ( \pmp k ) + i \lp [ k , \lmp ]
 \Big).
\end{equation}
As before, we use integration by parts to move the derivatives $\pp$ and $\pmp$ so they act on $\lap$, $\lamp$
\begin{equation}\label{s_fgt_7}
 \text{tr} \int\di^3x \di\theta^+ \pmm \Big(
 k \left( - \pp \lmp + \pmp \lp - 2i [ \lp , \lmp ] \right)
 \Big).
\end{equation}
This could be further simplified with the identity \eqref{s_kd}, and the result is 
\begin{equation}\label{s_fgt_lhs2}
 \text{tr} \int\di^3x \di\theta^+ \pmm \Big(
 - i k [ \lp , \lmp ]
 \Big).
\end{equation}
The sum of \eqref{s_fgt_lhs1} and \eqref{s_fgt_lhs2} gives the left side of equation \eqref{s_fgt_eq} and this
is equal to the right side, which is equal to a sum of \eqref{s_igt_bulk} and \eqref{s_igt_boundary}.
This, together with fulfillment of the boundary condition $\Sdx{}[\gamma_A;0,0]=0$, 
proves that \eqref{s_delta} correctly describes change of the Chern-Simons action for finite gauge transformations.

\subsection{Boundary Superfield}\label{sec_bs}

The Chern-Simons action is not gauge invariant, gauge transformations yield a contribution that does not vanish 
because of presence of the boundary. 
The gauge invariance can be restored if we assume that the apart from the bulk action given by 
\eqref{s_cs}, there is a boundary action that couples the gauge field to new boundary degrees of freedom.  
This boundary action has to possess the property that its  gauge transformation  
cancels the boundary terms \eqref{s_igt_bulk} and \eqref{s_igt_boundary} that were obtained from the gauge transformation 
of the Chern-Simons action \eqref{s_cs}.

The boundary part will not depend only on the gauge superfield but also on the scalar Lie-algebra valued 
boundary superfield $V$. For now, we will assume that the superfield $V$ is defined everywhere, 
we will see later that it suffices to define it on the boundary.
The gauge transformation of this superfield is postulated to be
\begin{equation}\label{l_fgt_v}
 e^{iV} \rightarrow e^{iV} e^{-iK}.
\end{equation}
It is chosen in this way in order to ensure that the connections $\Gamma_A^{(V)}$, which are finite gauge 
transformations of $\Gamma_A$ generated by $V$, are not changed by gauge transformations 
$\delta_g^{(K)}\Gamma_A^{(V)}=0$.
With the help of $\Gamma_A^{(V)}$ we can write the gauge invariant action as $\Sxx{cs}{}[\Gamma_A^{(V)}]$. 
The invariance of this action follows from the fact that gauge transformations leave $\Gamma_A^{(V)}$ unchanged.
The same procedure was also used in \cite{5} for a boundary in a space-like direction.

In the $SIM(1)$ setting we define two superfields corresponding to the Lorentz superfield $V$
\begin{align}\label{s_vproj}
 v &= V \vert_{\theta_+=0},
&
 \vm &= ( D_- V ) \vert_{\theta_+=0}.
\end{align}
The gauge invariant action can be written, according to \eqref{s_fgt_cs} as
\begin{multline}\label{s_vcs}
 \Sxx{cs}{} [ \gamma_A^{(V)} ] 
 = \Sxx{cs}{} [ \gamma_A ] + \Sdx{} [ \gamma_A ; v , \vm ]
 \\
 = \Sxx{cs}{bulk} [ \gamma_A ] + \left( \Sxx{cs}{boundary} [ \gamma_A ] + \Sdx{} [ \gamma_A ; v , \vm ] \right),
\end{multline}
where the expression inside brackets contains all boundary terms.
It is important to note that there is no dependence on the superfield $v$, $\vm$ in the bulk action,
only the surface action $\Sdx{}[\gamma_A;v,\vm]$ \eqref{s_delta} depends on these superfields. 
Furthermore, in order to evaluate the surface action we need to know only the value of $v$, $\vm$ and of their 
derivatives in directions tangent to the boundary. We do not need to know the derivatives $\pmm v$ or $\pmm\vm$ 
in direction normal to the boundary to evaluate the boundary action. Thus, it is enough if the superfields 
$v$, $\vm$ are defined on the boundary.

In the same way as we defined the superfields $\li{A}$, $\lai{A}$ we define the superfields $\ti{A}$, $\tai{A}$
\begin{align}\label{s_vdef}
 \tai{A} &= i ( \mathcal{D}_A e^{-is v} ) e^{is v},
&
 \ti{A} &= \tai{A} \vert_{s=1},
&
 \tm = i ( D_- e^{-is V} ) e^{is V} \vert_{\theta_+=0}.
\end{align}
Their infinitesimal gauge transformations are
\begin{align}\label{s_fgt_v}
 \delta_g^{(K)} \ti{A} &= i [ k, \ti{A} ] - \mathcal{D}_A k,
&
 \delta_g^{(K)} \tm &= i [ k, \tm ] - \km,
\end{align}
The superfields $\ti{A}$ and $\tai{A}$ satisfy the same set of identities \eqref{s_kda}, \eqref{s_kdpa},
\eqref{s_kalpha} and \eqref{s_kd} as $\li{A}$ and $\lai{A}$ with $v$ in place of $k$.
With these definitions we write the boundary part of the action as
\begin{multline}\label{s_vboundary}
 \Sdx{} [ \gamma_A ; v , \vm ] + \Sxx{cs}{boundary} [ \gamma_A ] = \frac{k}{4\pi} \text{tr} \int\di^3x \di\theta^+ \pmm \Big[ 
 - \frac{i}{6} ( \gm + \tm ) \{ \gp + \tp , \gp + \tp \}
 \\
 + \tp \gmp 
 - \gp \tmp
 + \int_0^1 \di s \left(  
    \left( \tfrac{\di}{\di s} \tap \right) \tamp 
  - \tap \left( \tfrac{\di}{\di s} \tamp \right)
 \right)
 \Big].
\end{multline}

There is an interesting interpretation for the combination $\gamma_A+\ti{A}$ that appears in the first term. 
If we replace the ordinary derivatives in the definition of $\ti{A}$ \eqref{s_vdef} with the covariant ones, then we get
\begin{equation}
 v^\nabla_A 
 = i ( \nabla_A e^{-iv} ) e^{iv}
 = i ( \mathcal{D}_A e^{-iv} - i \gamma_A e^{-iv} ) e^{iv}
 = \ti{A} + \gamma_A.
\end{equation}
Thus, the first term in \eqref{s_vboundary} can be written as
$-\frac{i}{6} ( \tm^\nabla ) \{ \tp^\nabla , \tp^\nabla \}$. This term is gauge invariant,
we do not need it to restore the gauge invariance, but we need it for $SIM(1)$ invariance.

\section{Chern-Simons Theory with Redefined \texorpdfstring{$SIM(1)$}{SIM(1)} Superfields}

In this section, we are going to rewrite the results of the previous sections using $SIM(1)$ superfields that
have better $SIM(1)$ transformation properties.
The description of the gauge theory in the previous sections was given with the help of a set of superfields 
$\gp$, $\gm$, $\gpp$, $\gmp$, $\gmm$. In this section, we are going to use a different set of superfields 
consisting of $\gp$, $\gx$, $\gpp$, $\gxp$, $\gxx$. The superfields $\gp$, $\gpp$ are defined according to \eqref{s_def}
the redefined superfields are defined as \cite{vf2}
\begin{align}\label{x_gdef}
 \gx &= i \left( \pxu{\alpha} \Gamma_\alpha \right) \vert_{\theta_+=0}
 = \gm - \pxm \gp ,
\nonumber\\
 \gxp &= i \left( \pxu{\alpha} \Gamma_{\alpha+} \right) \vert_{\theta_+=0} 
 = \gmp - \pxm \gpp ,
\nonumber\\
 \gxx &= -\left( \pxu{\alpha}\pxu{\beta} \Gamma_{\alpha\beta} \right)\vert_{\theta_+=0} 
 = \gmm -2 \pxm \gmp + \pxm^2 \gpp.
\end{align}
where the operator $\pxm$ is
\begin{equation}\label{x_pxd}
 \partial_{\times\alpha} = \frac{\partial_{+\alpha}}{\ppp}
 \qquad \Leftrightarrow \qquad
 \partial_{\times+} = 1,
 \qquad
 \pxm = \frac{\pmp}{\ppp}.
\end{equation}
The difference between $SIM(1)$ projections that have been used in the previous section and redefined superfields
is that each carry different representations. The $SIM(1)$ projections carry a spinor representation (and its tensor products), 
while redefined superfields carry representations that we have on $\mathcal{S}_{\text{quotient}}$ and 
$\mathcal{S}_{\text{invariant}}$ (and their tensor products).
If $\times$ is treated as a new type of index, together with $+$ an $-$, then the $SIM(1)$ transformation of any object
could be determined by applying the rules
\begin{align}\label{x_igs}
 \delta_s \psi_+ &= A \psi_+ ,
&
 \delta_s \psi_- &= - A \psi_- + B \psi_+ ,
&
 \delta_s \psi_\times &= -A \psi_\times,
\end{align}
on each index. If some object has only $+$ and $\times$ indices, which is the case of redefined superfields, 
then its $SIM(1)$ transformation is especially simple, it can be written as
\begin{equation}\label{tsimc}
 \delta_s \psi_{+\cdots+\times\cdots\times} = 
 A \cdot (\text{\# of ``}+\text{'' indices minus \# of ``}\times\text{'' indices}) \cdot \psi_{+\cdots+\times\cdots\times}.
\end{equation}

The infinitesimal gauge transformations of redefined superfields are more complicated that the ones we encountered 
in the case of $SIM(1)$ projections
\begin{align}\label{x_igt_g}
 \delta_g^{(K)} \gx &= i[k,\gx] - \pxu{\alpha}[k,\pxd{\alpha}\gp] + \kx,
\nonumber\\
 \delta_g^{(K)} \gxp &= i[k,\gxp] - \pxu{\alpha}[k,\pxd{\alpha}\gpp],
\nonumber\\
 \delta_g^{(K)} \gxx &= i[k,\gxx] - 2\pxu{\alpha}[k,\pxd{\alpha}\gxp] -i \pxu{\alpha}\pxu{\beta}[k,\pxd{\alpha}\pxd{\beta}\gpp] + \pxx k,
\end{align}
where $\kx=\km - \pxm \pp k$.

As before, we write the Chern-Simons action $\Sxx{cs}{}$ as a sum of a bulk part $\pSxx{cs}{bulk}$
and a boundary part $\pSxx{cs}{boundary}$
\begin{equation}\label{x_cs}
 \Sxx{cs}{} [ \gamma_A ] = \pSxx{cs}{bulk} [ \gamma_A ] + \pSxx{cs}{boundary} [ \gamma_A ].
\end{equation}
The prime is used in order to distinguish $\pSxx{cs}{bulk}$, $\pSxx{cs}{boundary}$ from the actions
$\Sxx{cs}{bulk}$, $\Sxx{cs}{boundary}$. The split of $\Sxx{cs}{}$ into $\pSxx{cs}{bulk}$, $\pSxx{cs}{boundary}$
is not the same as the split into $\Sxx{cs}{bulk}$, $\Sxx{cs}{boundary}$. In fact, we have
\begin{equation}\label{x_boundary_diff}
 \pSxx{cs}{boundary} - \Sxx{cs}{boundary} = - \left( \pSxx{cs}{bulk} - \Sxx{cs}{bulk} \right)
 = \frac{k}{4\pi} \text{tr} \int\di^3x \di\theta^+ \pmm \Big( \frac{i}{6} ( \pxm \gp ) \{ \gp , \gp \} \Big).
\end{equation}
In order to find this result, we have to keep track of $\pmm$ surface terms in the calculation of
the action $\pSxx{cs}{bulk}$ from $\Sxx{cs}{bulk}$ in \cite{vf2}. The only place where such a surface term 
appears is the identity
\begin{multline}\label{rss1}
 \text{tr} \int\di^3x \di\theta^+ \Big(
 - (\pxm\gpp)\left(\pxx\gp\right) + \gp\left(\pxx\pxm\gpp\right)
 \Big)
 \\
 =
 \text{tr} \int\di^3x \di\theta^+ \Big(
 - \frac{2}{3} \left( \pxx \gp \right) \left( \pxu{\alpha} \{ \gp , \pxd{\alpha} \} \right)
 \\
 - \frac{2i}{3} \gp [ \gpp , \pxm^3 \gpp ]
 + \frac{i}{6} \pmm \left( ( \pxm \gp ) \{ \gp , \gp \} \right)
 \Big).
\end{multline}
When this identity was derived in \cite{vf2}, the surface term was neglected.
We are going to provide a brief description of the proof that keeps track of the mentioned surface term.
We make the substitutions $\pxx=\pmm-\ppp\pxm^2$ and $\gpp=-\pp\gp+\frac{i}{2}\{\gp,\gp\}$ on the right side of
\eqref{rss1}, which gives us
\begin{multline}\label{rss2}
 \text{tr} \int\di^3x \di\theta^+ \Big(
   ( \pxm \pp \gp ) ( \pmm \gp ) 
 - \gp ( \pmm \pxm \pp \gp )
 - ( \pxm \pp \gp ) ( \ppp \pxm^2 \gp ) 
 \\
 + \gp ( \ppp \pxm^3 \pp \gp)
 - \frac{i}{2} ( \pxm \{ \gp , \gp \} ) ( \pmm \gp )
 + \frac{i}{2} \gp ( \pmm \pxm \{ \gp , \gp \} )
 \\
 + \frac{i}{2} ( \pxm \{ \gp , \gp \} ) ( \ppp \pxm^2 \gp )
 - \frac{i}{2} \gp ( \ppp \pxm^3 \{ \gp , \gp \} )
 \Big).
\end{multline}
The first term cancels with the second term, the third term cancels with the fourth term, the rest can be written as
\begin{multline}\label{rss3}
 \text{tr} \int\di^3x \di\theta^+ \Big(
 - \frac{2i}{3} ( \pxm \{ \gp , \gp \} ) ( \pmm \gp )
 + \frac{i}{3} \gp ( \pmm \pxm \{ \gp , \gp \} )
 \\
 + i ( \ppp \pxm^3 \gp ) \{ \gp , \gp \}
 + \frac{i}{6} \pmm \left( ( \pxm \gp ) \{ \gp , \gp \} \right)
 \Big).
\end{multline}
Notice, that a $\pmm$ surface term appeared as a result of this procedure.  
The rest of the calculation does not give any other $\pmm$ surface term.
The expression \eqref{rss3} can be written as (details can be found in \cite{vf2})
\begin{multline}\label{rss4}
 \text{tr} \int\di^3x \di\theta^+ \Big(
 - \frac{2}{3} \left( ( \pmm - \ppp \pxm^2 ) \gp \right) \left( i \pxm \{ \gp , \gp \} - i \{ \gp, \pxm \gp \} \right)
 \\
 - \frac{2i}{3} ( \ppp \gp ) \{ \gp , \pxm^3 \gp \}
 + \frac{i}{6} \pmm \left( ( \pxm \gp ) \{ \gp , \gp \} \right)
 \Big).
\end{multline}
This is exactly the expression that is on the right side of \eqref{rss1}.

The bulk action has already been calculated in \cite{vf2}
\begin{multline}\label{x_bulk}
 \pSxx{cs}{bulk} [ \gamma_A ] = \frac{k}{4\pi} \text{tr} \int\di^3x \di\theta^+ \bigg(
  -2 \gxx \left( \pp \gxp \right)
  - \gxp \left( \pxx \gp \right) + \left( \pxx \gxp \right) \gp
\\
  - \frac{2}{3} \left( \pxx \gp \right) \left( \pxu{\alpha} \left\{ \gp , \pxd{\alpha} \gp \right\} \right)
  + 2i \gp \left[ \gxp, \gxx \right]
\\
  -2 \gp \left[ \pxu{\alpha} \gxp , \pxd{\alpha} \gxp \right]
  +2 \gp \left[ \pxu{\alpha} \gpp , \pxd{\alpha} \gxx \right]
\\
  + 2i \gp \left[ \pxu{\alpha} \pxu{\beta} \gpp , \pxd{\alpha} \pxd{\beta} \gxp \right]
  - \frac{1}{3} \gp \left[ \pxu{\alpha} \pxu{\beta} \pxu{\gamma} \gpp , \pxd{\alpha} \pxd{\beta} \pxd{\gamma} \gpp \right]
 \bigg).
\end{multline}
The boundary action is obtained by combining the expression for $\Sxx{cs}{boundary}$ \eqref{s_boundary} 
with \eqref{x_boundary_diff}, it is given by 
\begin{equation}\label{x_boundary}
 \pSxx{cs}{boundary} [ \gamma_A ] = \frac{k}{4\pi} \text{tr} \int\di^3x \di\theta^+ \pmm \Big(
 - \frac{i}{6} \gx \{ \gp , \gp \}
 \Big).
\end{equation}

In this case both the bulk action $\pSxx{cs}{bulk}$ and the boundary action $\pSxx{cs}{boundary}$ are $SIM(1)$ invariant.
This contrasts with the case of the action \eqref{s_cs} where the change of the bulk action $\Sxx{cs}{bulk}$ 
had to be compensated by the change of the boundary action $\Sxx{cs}{boundary}$.

\subsection{Finite Gauge Transformation}

We are going to rewrite the expression for the surface term $\Sdx{}$ \eqref{s_delta} in terms of redefined superfields. Apart from the dependence on the superfields $\gp$, $\gx$, $\gpp$, $\gxp$, $\gxx$,
which have already been described, there will be a dependence on $\lp$, $\lx$, $\lxp$, $\lap$, $\laxp$, where
\begin{align}\label{x_kdef}
 \lx &= \lm - \pxm \lp,
&
 \lxp &= \lmp - \pxm \lpp,
&
 \laxp &= \lamp - \pxm \lapp.
\end{align}

We are going to make the following substitutions
\begin{align}
 \gm &= \gx + \pxm \gp,
&
 \gmp &= \gxp + \pxm \gpp
\nonumber\\
 \lm &= \lx + \pxm \lp,
&
 \lmp &= \lxp + \pxm \lpp,
&
 \lamp &= \laxp + \pxm \lapp.
\end{align}
The first term in \eqref{s_delta} gives
\begin{multline}\label{x_ap1}
 \text{tr} \int\di^3x \di\theta^+ \pmm \Big[
 - \frac{i}{6} ( \gm + \lm ) \{ \gp + \lp , \gp + \lp \}
 \Big]
 \\
 = 
 \text{tr} \int\di^3x \di\theta^+ \pmm \Big[
 - \frac{i}{6} ( \gx + \lx ) \{ \gp + \lp , \gp + \lp \}
 - \frac{i}{6} ( \pxm \gp ) \{ \gp , \gp \}
 \\
 - \frac{i}{6} ( \pxm \lp ) \{ \gp , \gp \} 
 - \frac{i}{3} \lp \{ \gp , \pxm \gp \}
 - \frac{i}{6} ( \pxm \gp ) \{ \lp , \lp \} 
 \\
 - \frac{i}{6} \gp \{ \lp , \pxm \lp \}
 - \frac{i}{6} ( \pxm \lp ) \{ \lp , \lp \}
 \Big],
\end{multline}
the second term gives
\begin{multline}\label{x_ap2}
 \text{tr} \int\di^3x \di\theta^+ \pmm \Big[
 \frac{i}{6} \gm \{ \gp , \gp \}
 \Big]
 \\
 = 
 \text{tr} \int\di^3x \di\theta^+ \pmm \Big[
   \frac{i}{6} \gx \{ \gp , \gp \}
 + \frac{i}{6} ( \pxm \gp ) \{ \gp , \gp \}
 \Big].
\end{multline}
The third and the fourth term give
\begin{multline}\label{x_ap3}
 \text{tr} \int\di^3x \di\theta^+ \pmm \Big[
   \lp \gmp 
 - \gp \lmp
 \Big]
 \\
 = 
 \text{tr} \int\di^3x \di\theta^+ \pmm \Big[
   \lp \gxp 
 - \gp \lxp
 + \lp ( \pxm \gpp )
 - \gp ( \pxm \lpp )
 \Big].
\end{multline}
It is convenient to rewrite this expression in such a way that there is no dependence on $\gpp$ and $\lpp$.
We can use \eqref{s_constraint} and \eqref{s_kdpa} (with $s=1$) to express $\gpp$ and $\lpp$ by expressions 
that contain only $\gp$, $\lp$. The result is
\begin{multline}\label{x_ap4}
 \text{tr} \int\di^3x \di\theta^+ \pmm \Big[
   \lp \gxp 
 - \gp \lxp
 - \lp ( \pxm \pp \gp )
 + \frac{i}{2} \lp ( \pxm \{ \gp, \gp \} )
 \\
 + \gp ( \pxm \pp \lp )
 + \frac{i}{2} \gp ( \pxm \{ \lp, \lp \} )
 \Big]
 \\
 = 
 \text{tr} \int\di^3x \di\theta^+ \pmm \Big[
   \lp \gxp 
 - \gp \lxp
 + \frac{i}{2} ( \pxm \lp ) \{ \gp, \gp \}
 + \frac{i}{2} ( \pxm \gp ) \{ \lp, \lp \}
 \Big],
\end{multline}
where in the equality we used the fact that if we integrate the third term $-\lp(\pxm\pp\gp)$ by parts 
to move $\pxm$ and $\pp$ we get $-\gp(\pxm\pp\lp)$, which cancels the fifth term.
The terms with the integral over $s$ in \eqref{s_delta} give
\begin{multline}\label{x_ap5}
 \text{tr} \int\di^3x \di\theta^+ \pmm \int_0^1 \di s \Big(  
   \left( \tfrac{\di}{\di s} \lap \right) \lamp 
 - \lap \left( \tfrac{\di}{\di s} \lamp \right)
 \Big)
 \\
 =
 \text{tr} \int\di^3x \di\theta^+ \pmm \int_0^1 \di s \Big(  
   \left( \tfrac{\di}{\di s} \lap \right) \laxp 
 - \lap \left( \tfrac{\di}{\di s} \laxp \right)
 \\
 + \left( \tfrac{\di}{\di s} \lap \right) ( \pxm \lapp ) 
 - \lap \left( \tfrac{\di}{\di s} \pxm \lapp \right)
 \Big).
\end{multline}
With the help of \eqref{s_kdpa}, we can write the last two terms only using $\lap$
\begin{multline}\label{x_ap6}
 \text{tr} \int\di^3x \di\theta^+ \pmm \int_0^1 \di s \Big(  
 - \left( \tfrac{\di}{\di s} \lap \right) ( \pxm \pp \lap ) 
 - \frac{i}{2} \left( \tfrac{\di}{\di s} \pxm \lap \right) \{ \lap , \lap \} 
 \\
 + \lap \left( \tfrac{\di}{\di s} \pxm \pp \lap \right)
 + \frac{i}{2} \lap \left( \tfrac{\di}{\di s} \pxm \{ \lap , \lap \} \right)
 \Big).
\end{multline}
The first and the third term vanish because they can be written as a total $\pp$ derivative.
The second and the fourth term can be written as
\begin{multline}\label{x_ap7}
 \text{tr} \int\di^3x \di\theta^+ \pmm \int_0^1 \di s \Bigg(
 \frac{i}{6} \frac{\di}{\di s} \left( (\pxm \lap) \{ \lap , \lap \} \right)
 \\
 - \frac{2i}{3} \left( \pxm \tfrac{\di}{\di s} \lap \right) \{ \lap , \lap \} 
 + \frac{2i}{3} \left( \tfrac{\di}{\di s} \lap \right) \{ \lap , \pxm \lap \} 
 \Bigg),
\end{multline}
and this can be written as  
\begin{multline}\label{x_ap8}
 \text{tr} \int\di^3x \di\theta^+ \pmm \Bigg(
 \frac{i}{6} (\pxm \lp) \{ \lp , \lp \}
 \\
 + \int_0^1 \di s \left[
 - \frac{2}{3} \left( \pxu{\alpha} \tfrac{\di}{\di s} \lap \right) \{ \lap , \pxd{\alpha} \lap \} 
 \right]
 \Bigg).
\end{multline}
Now, we are going to put together the pieces that we have just calculated. 
The sum of \eqref{x_ap1}, \eqref{x_ap2}, \eqref{x_ap4}, \eqref{x_ap8} gives us the action $\Sdx{}$
written with the help of redefined superfields
\begin{multline}\label{x_delta}
 \pSdx{} [ \gamma_A ; k, \kx ] = \frac{k}{4\pi} \text{tr} \int\di^3x \di\theta^+ \pmm \bigg(
 - \frac{i}{6} (\gx + \lx ) \{ \gp + \lp, \gp + \lp \}
 + \frac{i}{6} \gx \{ \gp , \gp \}
 \\
 + \lp \gxp 
 - \gp \lxp 
 + \frac{1}{3} ( \pxu{\alpha} \lp ) \{ \gp , \pxd{\alpha} \gp \} )
 + \frac{1}{3} ( \pxu{\alpha} \gp ) \{ \lp , \pxd{\alpha} \lp \} )
 \\
 + \int_0^1 \di s \Big(
   \left( \tfrac{\di}{\di s} \lap \right) \laxp
 - \lap \left( \tfrac{\di}{\di s} \laxp \right)
 - \frac{2}{3} \left( \pxu{\alpha} \tfrac{\di}{\di s} \lap \right) \{ \lap, \pxd{\alpha} \lap \}
 \Big)
 \bigg).
\end{multline}

\subsection{Boundary Superfield}

In section \ref{sec_bs}, we have coupled the gauge superfield to a boundary superfield in order to
restore the gauge invariance of the Chern-Simons theory. 
Now, we are going to reformulate this result with the help of redefined superfields.
The gauge invariant action can be written as
\begin{equation}\label{s_rvcs}
 \Sxx{cs}{} [ \gamma_A^{(V)} ] 
 = \pSxx{cs}{bulk} [ \gamma_A ] + \left( \pSxx{cs}{boundary} [ \gamma_A ] + \pSdx{} [ \gamma_A ; v , \vx ] \right),
\end{equation}
where the terms in brackets constitute the boundary part of the action. We should note that both the bulk part
and the boundary part are separately $SIM(1)$ invariant. 
In fact, each of actions $\pSxx{cs}{bulk}$, $\pSxx{cs}{boundary}$ and $\pSdx{}$ is separately invariant.
This contrasts with the case \eqref{s_vcs}, where $SIM(1)$ projections were used instead of redefined superfields.
In that case the parts $\Sxx{cs}{bulk}$ and $\Sxx{cs}{boundary}$ were not separately invariant.

The boundary part of the action from \eqref{s_rvcs} is
\begin{multline}\label{x_vboundary}
 \pSdx{} [ \gamma_A ; v, \vx ] + \pSxx{cs}{boundary} [ \gamma_A ] = 
 \frac{k}{4\pi} \text{tr} \int\di^3x \di\theta^+ \pmm \bigg(
 - \frac{i}{6} (\gx + \tx ) \{ \gp + \tp, \gp + \tp \}
 \\
 + \tp \gxp 
 - \gp \txp 
 + \frac{1}{3} ( \pxu{\alpha} \tp ) \{ \gp , \pxd{\alpha} \gp \} )
 + \frac{1}{3} ( \pxu{\alpha} \gp ) \{ \tp , \pxd{\alpha} \tp \} )
 \\
 + \int_0^1 \di s \Big(
   \left( \tfrac{\di}{\di s} \tap \right) \taxp
 - \tap \left( \tfrac{\di}{\di s} \taxp \right)
 - \frac{2}{3} \left( \pxu{\alpha} \tfrac{\di}{\di s} \tap \right) \{ \tap, \pxd{\alpha} \tap \}
 \Big)
 \bigg),
\end{multline}
where
\begin{align}\label{x_vdef}
 \tx &= \tm - \pxm \tp,
&
 \txp &= \tmp - \pxm \tpp,
&
 \taxp &= \tamp - \pxm \tapp.
\end{align}

\section{Conclusion}
In this paper, we have   analyzed a three dimensional supersymmetric Chern-Simons theory in presence of 
a boundary. This was done by 
  considering a boundary  that satisfied the condition $n\cdot x=0$, where
$n$ is a light-like vector. This boundary was called a light like boundary, and 
unlike the space-like boundary whose metric  was   rank two, the metric on this  boundary only was   rank one. 
The presence of this  boundary broke the symmetry group of the spacetime manifold from 
the Lorentz  group down to the $SIM(1)$ group. 
Thus, the  theory was studied using the $SIM(1)$ superspace. It was demonstrated that this theory only preserved 
half the supersymmetry of the original theory.   As the Chern-Simons theory 
had $\mathcal{N} =1$ supersymmetry in absence of a boundary, it only retained  
 $\mathcal{N}=1/2$ supersymmetry in presence of this boundary.  Finally, it was 
 observed that the Chern-Simons  theory can  be made gauge invariant by introducing new degrees of freedom 
 on the boundary. The gauge transformation of these new degrees of freedom exactly canceled the 
 boundary term obtained from the gauge transformation of the Chern-Simons theory. 

The results obtained in this paper could be used to study a system of multiple M2-branes in presence of a boundary in a light-like direction. This would require the coupling of matter fields in the bi-fundamental representation to the Chern-Simons theories. It may be noted that the coupling of matter fields to Yang-Mills theories has already been studied in fundamental representation \cite{vf1}. Furthermore, it would also be interesting to generalize this work by considering a  
Chern-Simons theory  in $\mathcal{N} =2$ superspace 
formalism. We can analyze the effect of imposing a boundary in the light-like direction for this Chern-Simons theory.   It is expected that half of the supersymmetry of the original Chern-Simons theory with $\mathcal{N} =2$ supersymmetry will be broken in the presence of a boundary.
Furthermore, it should also be possible to couple this theory to new degrees of freedom on the boundary such that the resultant theory is gauge invariant. 
The supersymmetry for an abelian   ABJM theory, in presence of a boundary, 
had also   discussed in $\mathcal{N} =2$ superspace formalism.
However, no discussion of supersymmetry of the full 
non-abelian ABJM theory in $\mathcal{N} =2$ superspace formalism, 
or its  gauge invariance had been done so far. 
So, a  generalization of these results to $\mathcal{N} =2$ superspace formalism,  and their 
application to   the ABJM theory in presence of a boundary will be interesting.  
As M2-branes can end on M5-branes, M9-branes or gravitational waves \cite{m9m2}, this formalism might be useful to study the physics of such systems.   

In order to quantize the action for multiple M2-branes we have to add a gauge fixing term and a ghost term to the original action.
The total action thus obtained will be invariant under BRST symmetry \cite{brst1}-\cite{qbrst1}. 
The BRST symmetry   for multiple M2-branes  on a manifold without a boundary
has  been  studied in $\mathcal{N} =1$ superspace formalism
\cite{mf012}-\cite{mf01}. This analysis has been generalized to include a   boundary in a space-like direction   \cite{5}. 
It has been demonstrated that the bulk action for multiple M2-branes  is not invariant under the 
BRST  transformations. 
The BRST transformations for this action generate a boundary contribution. 
However, the BRST transformations of the new boundary degrees of freedom   exactly cancel the 
boundary contribution generated from the BRST transformation of the bulk action for M2-branes. 
It will be interesting to investigate this for a boundary in the light-like direction.   

It has been demonstrated that using the Ho\v{r}ava-Witten theory,  
one of the low energy limits of the heterotic string theory can be obtained from the  
eleven dimensional supergravity in presence of a  boundary   \cite{hwsetup}-\cite{hwsetup1}. Thus,
  the strong-coupling limit of the type IIA string theory has been related to 
the strong-coupling limit of the heterotic string \cite{az5}-\cite{az4}. 
This was done by compactifying the original theory on 
an interval bounded by mirror  orientifold  planes. It was argued that a ten dimensional 
$E_8$ super-Yang-Mills theory appears on each plane. So, 
two $E_8$ gauge theories were obtained  on the mirror planes, and supergravity  was obtained between these planes. 
In this construction, the low-energy value of the Newton's constant decreases when the distance between  the planes is increased. The gauge coupling remains
fixed as this interval is increased. Thus, by adjustment of the length of this interval, 
it was possible to obtain  a unification of gauge and gravitational couplings. In this theory, 
six   dimensions were  compactified, and thus, a five dimensional theory on an interval
with mirror plane boundaries was  obtained. This theory is expected to be a  five dimensional 
supergravity model, with additional bulk super-multiplets. It has been argued that the   
analysis of a simpler system can help understand   the  Ho\v{r}ava-Witten theory. 
A simplified construction of a five dimensional globally supersymmetric Yang-Mills theory coupled to a
four dimensional hypermultiplet   on the boundary has also been analyzed \cite{az2}.
It would be interesting to analyze what features 
of this model can be retained for a boundary in a light like direction.
Thus, it will be interesting to generalize the results of this paper to five dimensions and use it for analyzing a globally supersymmetric Yang-Mills theory.


\begin{thebibliography}{99}
 
\bibitem{1}
  S.~J.~Gates, M.~T.~Grisaru, M.~Rocek and W.~Siegel,
  ``Superspace Or One Thousand and One Lessons in Supersymmetry,''
  Front.\ Phys.\  {\bf 58}, 1 (1983)
  [hep-th/0108200].

\bibitem{a}
  D.~V.~Belyaev,
  ``Boundary conditions in the Mirabelli and Peskin model,''
  JHEP {\bf 0601}, 046 (2006)
  [hep-th/0509171].

\bibitem{b}
  D.~V.~Belyaev,
  ``Boundary conditions in supergravity on a manifold with boundary,''
  JHEP {\bf 0601}, 047 (2006)
  [hep-th/0509172].

\bibitem{c}
  P.~van Nieuwenhuizen and D.~V.~Vassilevich,
  ``Consistent boundary conditions for supergravity,''
  Class.\ Quant.\ Grav.\  {\bf 22}, 5029 (2005)
  [hep-th/0507172].

\bibitem{d}
  U.~Lindstrom, M.~Rocek and P.~van Nieuwenhuizen,
  ``Consistent boundary conditions for open strings,''
  Nucl.\ Phys.\ B {\bf 662}, 147 (2003)
  [hep-th/0211266].

\bibitem{e}
  P.~Di Vecchia, B.~Durhuus, P.~Olesen and J.~L.~Petersen,
  ``Fermionic Strings With Boundary Terms,''
  Nucl.\ Phys.\ B {\bf 207}, 77 (1982).

\bibitem{f}
  P.~Di Vecchia, B.~Durhuus, P.~Olesen and J.~L.~Petersen,
  ``Fermionic Strings With Boundary Terms. 2. The O(2) String,''
  Nucl.\ Phys.\ B {\bf 217}, 395 (1983).

\bibitem{g}
  Y.~Igarashi,
  ``Supersymmetry and the Casimir Effect Between Plates,''
  Phys.\ Rev.\ D {\bf 30}, 1812 (1984).

\bibitem{1z}
  D.~V.~Belyaev and P.~van Nieuwenhuizen,
  ``Rigid supersymmetry with boundaries,''
  JHEP {\bf 0804}, 008 (2008)
  [arXiv:0801.2377 [hep-th]].

\bibitem{8}
  M.~Faizal and D.~J.~Smith,
  ``Nonanticommutativity in the presence of a boundary,''
  Phys.\ Rev.\ D {\bf 87}, no. 2, 025019 (2013)
  [arXiv:1211.3654 [hep-th]].

\bibitem{9}
  M.~Faizal,
  ``Deformed Super-Yang-Mills in Batalin-Vilkovisky Formalism,''
  Int.\ J.\ Theor.\ Phys.\  {\bf 52}, 392 (2013)
  [arXiv:1209.2357 [hep-th]].

\bibitem{4z}
  D.~S.~Berman and D.~C.~Thompson,
  ``Membranes with a boundary,''
  Nucl.\ Phys.\ B {\bf 820}, 503 (2009)
  [arXiv:0904.0241 [hep-th]].
  
\bibitem{abjm}
  O.~Aharony, O.~Bergman, D.~L.~Jafferis and J.~Maldacena,
  ``N=6 superconformal Chern-Simons-matter theories, M2-branes and their gravity duals,''
  JHEP {\bf 0810}, 091 (2008)
  [arXiv:0806.1218 [hep-th]].

\bibitem{ab}
  M.~A.~Bandres, A.~E.~Lipstein and J.~H.~Schwarz,
  ``Studies of the ABJM Theory in a Formulation with Manifest SU(4) R-Symmetry,''
  JHEP {\bf 0809}, 027 (2008)
  [arXiv:0807.0880 [hep-th]].

\bibitem{ab1}
  M.~Schnabl and Y.~Tachikawa,
  ``Classification of N=6 superconformal theories of ABJM type,''
  JHEP {\bf 1009}, 103 (2010)
  [arXiv:0807.1102 [hep-th]].

\bibitem{ab2}
  E.~Antonyan and A.~A.~Tseytlin,
  ``On 3d N=8 Lorentzian BLG theory as a scaling limit of 3d superconformal N=6 ABJM theory,''
  Phys.\ Rev.\ D {\bf 79}, 046002 (2009)
  [arXiv:0811.1540 [hep-th]].

\bibitem{abjm2}
  O.~K.~Kwon, P.~Oh and J.~Sohn,
  ``Notes on Supersymmetry Enhancement of ABJM Theory,''
  JHEP {\bf 0908}, 093 (2009)
  [arXiv:0906.4333 [hep-th]].
  
\bibitem{2abjm}
  H.~Samtleben and R.~Wimmer,
  ``N=6 Superspace Constraints, SUSY Enhancement and Monopole Operators,''
  JHEP {\bf 1010}, 080 (2010)
  [arXiv:1008.2739 [hep-th]].

\bibitem{1abcd}
  A.~Gustavsson,
  ``Selfdual strings and loop space Nahm equations,''
  JHEP {\bf 0804}, 083 (2008)
  [arXiv:0802.3456 [hep-th]].

\bibitem{2abcd}
  J.~Bagger and N.~Lambert,
  ``Comments on multiple M2-branes,''
  JHEP {\bf 0802}, 105 (2008)
  [arXiv:0712.3738 [hep-th]].

\bibitem{3abcd}
  J.~Bagger and N.~Lambert,
  ``Gauge symmetry and supersymmetry of multiple M2-branes,''
  Phys.\ Rev.\ D {\bf 77}, 065008 (2008)
  [arXiv:0711.0955 [hep-th]].

\bibitem{5}  M.~Faizal and D.~J.~Smith,
  ``Supersymmetric Chern-Simons Theory in Presence of a Boundary,''
  Phys.\ Rev.\ D {\bf 85} (2012) 105007
  [arXiv:1112.6070 [hep-th]].

\bibitem{5a}
  M.~Faizal,
  ``Boundary Effects in the BLG Theory,''
  Mod.\ Phys.\ Lett.\ A {\bf 29}, no. 31, 1450154 (2014)
  [arXiv:1303.5477 [hep-th]].

\bibitem{6}
  M.~Faizal,
  ``Gauge and Supersymmetric Invariance of a Boundary Bagger-Lambert-Gustavsson Theory,''
  JHEP {\bf 1204}, 017 (2012)
  [arXiv:1204.0297 [hep-th]].
  
  \bibitem{d4}
  D.~J.~Smith,
  ``Intersecting brane solutions in string and M theory,''
  Class.\ Quant.\ Grav.\  {\bf 20}, R233 (2003)
  [hep-th/0210157].

\bibitem{1d}
  H.~Nastase, C.~Papageorgakis and S.~Ramgoolam,
  ``The Fuzzy S**2 structure of M2-M5 systems in ABJM membrane theories,''
  JHEP {\bf 0905}, 123 (2009)
  [arXiv:0903.3966 [hep-th]].

\bibitem{5d}
  J.~Armas and M.~Blau,
  ``Black probes of Schrödinger spacetimes,''
  JHEP {\bf 1408}, 140 (2014)
  [arXiv:1405.1301 [hep-th]].

\bibitem{6d}
  J.~Armas and T.~Harmark,
  ``Constraints on the effective fluid theory of stationary branes,''
  JHEP {\bf 1410}, 63 (2014)
  [arXiv:1406.7813 [hep-th]].

\bibitem{2d}
  R.~Iengo and J.~G.~Russo,
  ``Non-linear theory for multiple M2 branes,''
  JHEP {\bf 0810}, 030 (2008)
  [arXiv:0808.2473 [hep-th]].

\bibitem{d5}
  D.~Youm,
  ``Partially localized intersecting BPS branes,''
  Nucl.\ Phys.\ B {\bf 556}, 222 (1999).

\bibitem{hwsetup}
  P.~Horava and E.~Witten,
  ``Heterotic and type I string dynamics from eleven-dimensions,''
  Nucl.\ Phys.\ B {\bf 460}, 506 (1996)
  [hep-th/9510209].

\bibitem{hwsetup1}
  P.~Horava and E.~Witten,
  ``Eleven-dimensional supergravity on a manifold with boundary,''
  Nucl.\ Phys.\ B {\bf 475}, 94 (1996)
  [hep-th/9603142].

\bibitem{d12}
  C.~S.~Chu and D.~J.~Smith,
  ``Towards the Quantum Geometry of the M5-brane in a Constant C-Field from Multiple Membranes,''
  JHEP {\bf 0904}, 097 (2009)
  [arXiv:0901.1847 [hep-th]].

\bibitem{M5BLG}
  P.~M.~Ho,
  ``A Concise Review on M5-brane in Large C-Field Background,''
  Chin.\ J.\ Phys.\  {\bf 48}, 1 (2010)
  [arXiv:0912.0445 [hep-th]].

\bibitem{f1}
  D.~N.~Kabat and W.~Taylor,
  ``Spherical membranes in matrix theory,''
  Adv.\ Theor.\ Math.\ Phys.\  {\bf 2}, 181 (1998)
  [hep-th/9711078].

\bibitem{f0}
  J.~Castelino, S.~Lee and W.~Taylor,
  ``Longitudinal five-branes as four spheres in matrix theory,''
  Nucl.\ Phys.\ B {\bf 526}, 334 (1998)
  [hep-th/9712105].

\bibitem{f5}
  N.~R.~Constable, R.~C.~Myers and O.~Tafjord,
  ``The Noncommutative bion core,''
  Phys.\ Rev.\ D {\bf 61}, 106009 (2000)
  [hep-th/9911136].

\bibitem{f4}
  N.~R.~Constable, R.~C.~Myers and O.~Tafjord,
  ``NonAbelian brane intersections,''
  JHEP {\bf 0106}, 023 (2001)
  [hep-th/0102080].

\bibitem{f2}
  P.~L.~H.~Cook, R.~de Mello Koch and J.~Murugan,
  ``NonAbelian bionic brane intersections,''
  Phys.\ Rev.\ D {\bf 68}, 126007 (2003)
  [hep-th/0306250].

\bibitem{amodel}
  A.~Brini,
  ``Open topological strings and integrable hierarchies: Remodeling the A-model,''
  Commun.\ Math.\ Phys.\  {\bf 312}, 735 (2012)
  [arXiv:1102.0281 [hep-th]].

\bibitem{bmodel}
  S.~Hyun, K.~Oh, J.~D.~Park and S.~H.~Yi,
  ``Topological B-model and c=1 string theory,''
  Nucl.\ Phys.\ B {\bf 729}, 135 (2005)
  [hep-th/0502075].

\bibitem{vf1}
  J.~Vohánka and M.~Faizal,
  ``Super-Yang-Mills Theory in SIM(1) Superspace,''
  Phys.\ Rev.\ D {\bf 91} (2015) 4,  045015
  [arXiv:1409.6334 [hep-th]].

\bibitem{v1}
  J.~Vohanka,
  ``Gauge Theory and SIM(2) Superspace,''
  Phys.\ Rev.\ D {\bf 85}, 105009 (2012)
  [arXiv:1112.1797 [hep-th]].

\bibitem{1v}
  S.~Petras, R.~von Unge and J.~Vohanka,
  ``SIM(2) and supergraphs,''
  JHEP {\bf 1107}, 015 (2011)
  [arXiv:1102.3856 [hep-th]].

\bibitem{vf2}
  J.~Vohánka and M.~Faizal,
  ``Chern-Simons Theory in SIM(1) Superspace,''
  arXiv:1503.04761 [hep-th].

\bibitem{m9m2}
  D.~S.~Berman, M.~J.~Perry, E.~Sezgin and D.~C.~Thompson,
  ``Boundary Conditions for Interacting Membranes,''
  JHEP {\bf 1004}, 025 (2010)
  [arXiv:0912.3504 [hep-th]].

\bibitem{brst1}
  C.~Becchi, A.~Rouet and R.~Stora,
  ``Renormalization of Gauge Theories,''
  Annals Phys.\  {\bf 98}, 287 (1976).

\bibitem{brst12}
  I.~V.~Tyutin,
  ``Gauge Invariance in Field Theory and Statistical Physics in Operator Formalism,''
  arXiv:0812.0580 [hep-th].

\bibitem{qbrst1}
  T.~Kugo and I.~Ojima,
  ``Local Covariant Operator Formalism of Nonabelian Gauge Theories and Quark Confinement Problem,''
  Prog.\ Theor.\ Phys.\ Suppl.\  {\bf 66}, 1 (1979).

\bibitem{mf012}
  M.~Faizal,
  ``M-Theory on Deformed Superspace,''
  Phys.\ Rev.\ D {\bf 84}, 106011 (2011)
  [arXiv:1111.0213 [hep-th]].
  
\bibitem{mf01}
  M.~Faizal,
  ``$M$-Theory in the Gaugeon Formalism,''
  Commun.\ Theor.\ Phys.\  {\bf 57}, 637 (2012)
  [arXiv:1201.1220 [hep-th]].

\bibitem{az5}
  E.~Witten,
  ``Strong coupling expansion of Calabi-Yau compactification,''
  Nucl.\ Phys.\ B {\bf 471}, 135 (1996)
  [hep-th/9602070].

\bibitem{az4}
  P.~Horava,
  ``Gluino condensation in strongly coupled heterotic string theory,''
  Phys.\ Rev.\ D {\bf 54}, 7561 (1996)
  [hep-th/9608019].

\bibitem{az2}
  E.~A.~Mirabelli and M.~E.~Peskin,
  ``Transmission of supersymmetry breaking from a four-dimensional boundary,''
  Phys.\ Rev.\ D {\bf 58}, 065002 (1998)
  [hep-th/9712214].

\end{thebibliography}
\end{document}